\documentclass[aip,jap,reprint,superscriptaddress,nopacs]{revtex4-1}

\usepackage{graphicx}
\usepackage{amsmath}
\usepackage{float}

\usepackage[driverfallback=dvipdfm, colorlinks=true,citecolor=blue,linkcolor=magenta]{hyperref}
\hypersetup{colorlinks = true, linkcolor = blue, urlcolor=blue, bookmarksnumbered =  true}

\begin{document}
\title{Site selective growth of heteroepitaxial diamond nanoislands containing single SiV centers}
\author{Carsten Arend}
\affiliation{FR 7.2 (Experimentalphysik), Universit\"at des Saarlandes, Campus E2.6, D-66123 Saarbr\"ucken}
\author{Patrick Appel}
\affiliation{Departement Physik, Universit\"at Basel, Klingelbergstrasse 82, CH-4056 Basel}
\author{Jonas Nils Becker}
\affiliation{FR 7.2 (Experimentalphysik), Universit\"at des Saarlandes, Campus E2.6, D-66123 Saarbr\"ucken}
\author{Marcel Schmidt}
\affiliation{FR 7.2 (Experimentalphysik), Universit\"at des Saarlandes, Campus E2.6, D-66123 Saarbr\"ucken}
\author{Martin Fischer}
\affiliation{Lehrstuhl f\"ur Experimentalphysik IV, Universit\"at Augsburg, D-86135 Augsburg}
\author{Stefan Gsell}
\affiliation{Lehrstuhl f\"ur Experimentalphysik IV, Universit\"at Augsburg, D-86135 Augsburg}
\author{Matthias Schreck}
\affiliation{Lehrstuhl f\"ur Experimentalphysik IV, Universit\"at Augsburg, D-86135 Augsburg}
\author{Christoph Becher}
\affiliation{FR 7.2 (Experimentalphysik), Universit\"at des Saarlandes, Campus E2.6, D-66123 Saarbr\"ucken}
\author{Patrick Maletinsky}
\affiliation{Departement Physik, Universit\"at Basel, Klingelbergstrasse 82, CH-4056 Basel}
\author{Elke Neu}
\affiliation{FR 7.2 (Experimentalphysik), Universit\"at des Saarlandes, Campus E2.6, D-66123 Saarbr\"ucken}
\affiliation{Departement Physik, Universit\"at Basel, Klingelbergstrasse 82, CH-4056 Basel}

\date{\today}

\begin{abstract}
We demonstrate the controlled preparation of heteroepitaxial diamond nano- and microstructures on silicon wafer based iridium films as hosts for single color centers. Our approach uses electron beam lithography followed by reactive ion etching to pattern the carbon layer formed by bias enhanced nucleation on the iridium surface. In the subsequent chemical vapor deposition process, the patterned areas evolve into regular arrays of (001) oriented diamond nano-islands with diameters of $<500~$nm and a height of $\approx 60$ nm. In the islands, we identify single SiV color centers with narrow zero phonon lines down to 1 nm at room temperature.
\end{abstract}

\maketitle

Color centers in diamond are being extensively investigated as stable, room temperature single photon emitters \cite{Neu2014b} simultaneously hosting highly controllable electronic spin systems. They are promising candidates for quantum information processing architectures (single photon sources and spin qubits) and as quantum sensors (e.g.\ for magnetic fields\cite{Rondin2014} and optical near fields \cite{Tisler2013a}). For all these applications, efficient collection of the fluorescence light emitted by color centers is crucial. The high refractive index of diamond ($n=2.4$) here is an ambivalent property: on one hand, it renders light extraction from bulk material highly challenging due to total internal reflection; on the other hand, it enables controlling the emission properties of color centers using versatile diamond photonic structures like waveguides and nanocavities.\cite{Aharonovich2014a} Recent work on color centers in diamond often uses top-down fabricated single-crystal diamond nano/microstructures enabling efficient light collection. Top-down nanofabrication creates many photonic structures, e.g.\ nanopillars, in regular arrays in which color centers can be straightforwardly identified and (re-)addressed.\cite{Babinec2010} However, diamond nanofabrication requires sophisticated, non-standard procedures e.g.\ for plasma etching, that are challenging due to the chemical inertness of diamond. Avoiding diamond nanofabrication,\cite{Babinec2010} enhanced out-coupling of light is alternatively obtained using nanodiamonds (NDs) combined with non-diamond photonic structures.\cite{Aharonovich2014a} However, color centers in NDs may suffer from unstable fluorescence and short spin coherence times. Random spatial placement and orientation of NDs, e.g.\ resulting from spin-coating deposition, renders identifying and re-addressing suitable color centers challenging. 
In this paper, we introduce an approach to unite the advantages of ND based systems and top-down fabrication of photonic structures i.e. controlled growth of regular arrays of heteroepitaxial diamond nano- and microstructures on iridium (Ir).

In previous work, regular arrays of diamond nanostructures have been obtained by chemical vapor deposition (CVD) on diamond substrates through openings in a mask.\cite{Aharonovich2013,Sovyk2014,Masuda2001,Furuyama2015} However, etching of the mask material in the CVD plasma led to the formation of high densities of color centers.\cite{Aharonovich2013,Sovyk2014} Thus, this method can so far not be considered as an approach capable of high purity diamond growth for single color center applications.  Moreover, for this system a significant fraction of color center fluorescence is emitted into the growth substrate (diamond).\cite{Aharonovich2013,Sovyk2014} Using ND seeding, growth of NDs or ND clusters at pre-defined positions on various substrates has been achieved.\cite{Shimoni2014,Singh2014} However, in all these experiments, identifying single color centers wasn't feasible. 

Here, we present an alternative approach to grow regular patterns of diamond micro- and nanostructures for single photon emission. It is based on diamond heteroepitaxy on iridium/yttria-stabilized-zirconia/silicon (Ir/YSZ/Si) substrates. This system has recently been suggested as a highly promising route towards large area single crystal diamond wafers.\cite{Gsell2004} Furthermore, the excellent inertness of Ir surfaces towards etching in the CVD growth environment facilitates the synthesis of high purity diamond crystals. Patterned diamond growth has been demonstrated earlier on Ir/MgO substrates,\cite{Ando2004,Ando2012} however, no investigations of color centers have been performed. In previous work, CVD-NDs grown on ND seeds hosted bright, single silicon vacancy (SiV) centers created \textit{in-situ} as a result of residual Si incorporation.\cite{Neu2011} The presence of the Ir substrates leads to potential collection efficiencies for fluorescence light exceeding 65\% (numerical aperture 0.8) via altering the emission characteristics of the color centers.\cite{Neu2012a} Heteroepitaxial (001) oriented NDs (nanoislands, NIs) on a randomly structured Ir substrate were used for the spectroscopy of single SiV centers\cite{Neu2011b} which showed MHz single photon-rates.\cite{Neu2012a} However, in all previous work, random spatial placement of NDs or NIs rendered it highly challenging to re-identify promising single SiV centers and use them for applications.   
         
In this work, we have grown arrays of diamond nano- and microstructures using patterned etching of the carbon layer formed by bias enhanced nucleation on Ir. The procedure involves electron beam lithography as well as a dry etch step in an inductively coupled plasma reactive ion etching system (ICP-RIE) and thus allows to engineer structures of almost arbitrary, planar geometries. Depending on the growth conditions, we achieve almost perfect selectivity between etched and non-etched areas. Silicon Vacancy (SiV) centers are created \textit{in-situ}; and we observe single centers in the diamond nanostructures. 
 
\begin{figure}
\includegraphics[width=8.5cm]{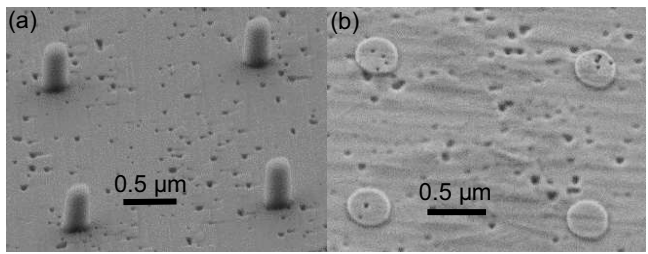}
\caption{Scanning electron microscopy (SEM) image under 45$^\circ$ of (a) example of cylindrical HSQ masks on Ir (b) example of circular nucleation areas defined using $60~$s of ICP-RIE plasma after HSQ mask removal before CVD growth of diamond. \label{Figmask} }
\end{figure}
As growth substrate, we employ an Ir/YSZ/Si multilayer system fabricated on Si(001) with an off-axis angle of 4$^\circ$ (for details see Refs.\ \onlinecite{Gsell2004,Fischer2008a}). Heteroepitaxial diamond nucleation is achieved by bias enhanced nucleation (BEN) using a bias voltage of  $\approx-300~$V in a gas atmosphere of 3\% CH$_4$ in H$_2$. The typical density of nuclei created in this process is $\approx 3\times10^{11}~$$\textrm{cm}^{-2}$. 
To define the patterns, hydrogen silsesquioxane (HSQ) negative electron beam resist (FOX-16, Dow Corning) is spun onto the top Ir layer. We use e-beam lithography ($30~$keV) to pattern etch-masks into the HSQ layer of $\approx300-600~$nm thickness [example of HSQ mask see Fig.\ \ref{Figmask}(a)]. After developing the mask, we etch away the BEN layer, which is highly chemically stable but only $\approx 1~$nm thick.\cite{Brescia2008} Consequently, the process also etches some of the underlying Ir to ensure full removal of the BEN layer.  We use an ICP-RIE (Sentech SI 500) reactor using the following plasma parameters: 50\% Ar and O$_2$ respectively, gas flow $50~$sccm each, pressure $0.5~$Pa, ICP power $500~$W, bias power $200~$W.  Ar ensures physical etching of BEN layer and Ir; whereas O$_2$ ensures full oxidation and thus removal of all carbon components. From scanning electron images of samples etched for  $60~$s [see Fig.\ \ref{Figmask}(b)], we estimate an etch rate of $\approx25~$nm/min for the Ir layer. Note that this short ICP-RIE etch step does not induce any discernible erosion of the HSQ mask,\cite{suppl2015} indicating that much thinner etch masks might be employed in future work.  Efficient etching of the BEN layer is confirmed by the observation (data not shown) that already $5~$s of etching remove the BEN layer completely and efficiently suppress heteroepitaxial growth of diamond. After etching, we remove the HSQ mask using buffered oxide etch and rinsing in deionized water. As visible in Fig.\ \ref{Figmask}, the Ir layer shows a residual roughness with slight dimples of typically tens of nanometers depth. We note that these dimples, which result from island formation in the nucleation stage of the Ir layer deposition, do not penetrate through the layer and do not negatively influence diamond heteroepitaxy.

\begin{figure}
\includegraphics[width=8.5cm]{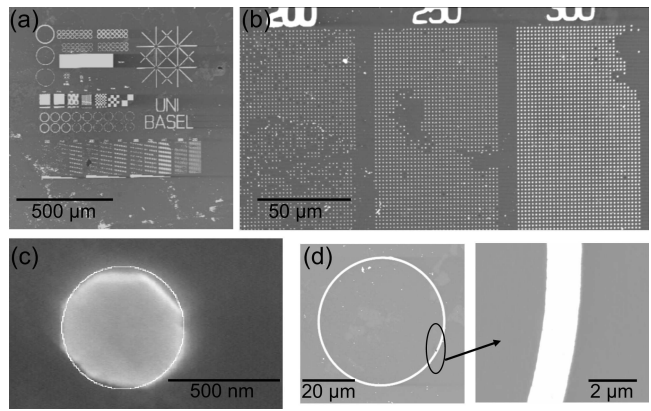}
\caption{SEM images (top view) of heteroepitaxial CVD diamond (a) Overview of the pattern (b) NIs with $<500~$nm diameter (c) close up of a diamond NI created by CVD (white circle as guide to the eye) (d) ring structure demonstrating very smooth diamond structures with a diamond thickness of $60~$nm. We conclude from additional SEM and AFM images, that the top surfaces of the NIs are flat and thus resemble (001) surfaces, whereas the edges of the structures closely follow the pre-defined growth areas. \label{FigCVD} }
\end{figure}
On the structured substrates, we grow heteroepitaxial diamond using a microwave-plasma-assisted CVD process (IPLAS reactor with CYRANNUS plasma source). We employ a CH$_4$/H$_2$ plasma containing 0.5 or 0.2\% CH$_4$ at a gas pressure of $30~$mbar and a microwave power of $2000~$W. For 0.5\% CH$_4$, we find a growth rate of $\approx150~$nm/h, while for 0.2\% CH$_4$ slower growth with only $\approx70~$nm/h is observed. SiV centers are created during CVD growth as a result of etching of the sidewalls of the underlying silicon substrate. Growth at 0.5\% CH$_4$ yields almost perfect selectivity while for 0.2\% CH$_4$ and longer growth time the selectivity decreases and non-epitaxial crystals appear. Figure \ref{FigCVD}(a) shows an overview of the pattern grown with 0.5\% CH$_4$ and high selectivity. To test the versatility of the method, we structure not only NIs but also some larger structures like rings, checkerboard patterns and the logo of Basel university. Figure \ref{FigCVD}(b) shows an array of circular NIs, while Fig.\ \ref{FigCVD}(c) shows a close up of a NI with $<500~$nm diameter which is closely resembling the circular shape of the growth area defined in the nanofabrication process. The NIs are especially suitable to observe single color centers due to a small diamond volume and thus higher probability to isolate single color centers. Some missing NIs in the regular pattern arise where the HSQ mask was lost due to insufficient adhesion which might be improved in the future using an adhesion promoter. The ring structures demonstrate the growth of smooth diamond films, which follow the predefined growth areas almost ideally [Fig.\ \ref{FigCVD}(d)]. Note some minor diamond growth outside the pre-defined areas, e.g.\ inside the diamond ring in Fig.\ \ref{FigCVD}(d). This arises most probably due to surface contamination that  protected the underlying BEN layer during etching.

The samples are analyzed using a homebuilt confocal fluorescence microscope with a Ti:sapphire laser at $700~$nm as excitation source. A 100x microscope objective with a numerical aperture of 0.8 focuses the laser light onto the sample and collimates the photoluminescence (PL). Excitation laser and PL are separated by a glass beam splitter and dielectric longpass and bandpass filters. A singlemode fiber serves as pinhole for the confocal setup and directs the PL to either a Hanbury-Brown Twiss setup (HBT) consisting of two avalanche photodiodes (APD, Perkin Elmer SPCM-AQRH-14) to obtain photon statistics and countrate or to a grating spectrometer (Horiba Jobin Yvon, iHR 550 and Symphony CCD) for spectral analysis. Gratings with 600 (1800) grooves per mm provide a resolution of $0.18 (0.06)~$nm. A liquid helium flow cryostat (Janis Research, ST-500LN) can be used for experiments at cryogenic temperatures.

\begin{figure}
\includegraphics[width=7.5cm]{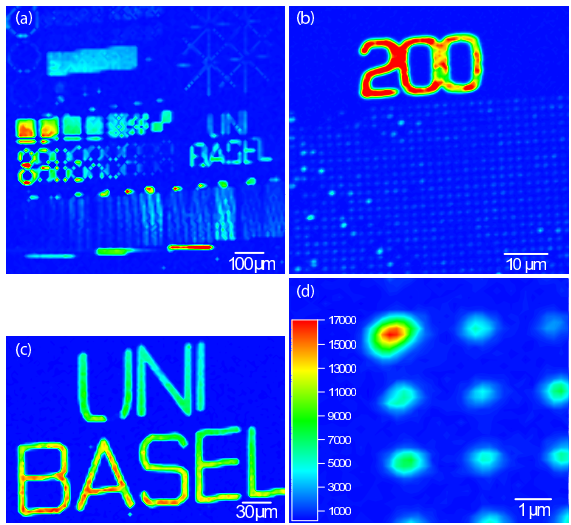}
\caption{(color online) PL image of (a) the complete pattern of heteroepitaxial CVD diamond (b) NIs with $<500~$nm diameter (c) the Basel university logo (d) spatially well separated NIs at $100~\mu$W excitation power.\label{FigScans} }
\end{figure}

\begin{figure}
\includegraphics[width=8.5cm]{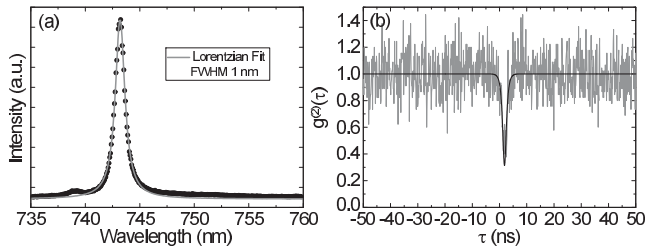}
\caption{(a) ZPL of an individual SiV center showing a linewidth of 1 nm (b) $g^{(2)}$ function of the emitter measured at low excitation power ($25~\mu$W) fitted with the model for a two level system\label{FigZPLg2}}
\end{figure}

Confocal images obtained recording the PL in a spectral window of 730-750 nm [see Fig.\ \ref{FigScans}] almost perfectly reproduce the diamond pattern fabricated earlier. Figure \ref{FigScans}(a) shows a coarse PL image of the pattern in Fig.\ \ref{FigCVD}(a), with close up images shown in Figs.\ \ref{FigScans}(b) and (c). We observe strong PL from the diamond pattern, while almost no (background) PL is emitted from the etched Ir areas. Examining the spatially well separated NIs [see Fig.\ \ref{FigScans}(d)] and other microstructures on the sample, localized bright spots emerge in the images, while the larger structures predominantly show uniform luminescence intensity see e.g. Fig.\ \ref{FigScans}(c).
Most of the bright spots exhibit spectra with broad ($> 4~$nm) zero phonon lines (ZPL) or multiple ZPLs as characteristic for SiV ensembles. Additionally, we find emitters with narrow linewidth SiV ZPLs localized around $740~$nm and linewidths from 1 to $1.8~$nm at room temperature [see Fig.\ \ref{FigZPLg2} and Ref.\ \onlinecite{suppl2015}]. As a rough order of magnitude for the statistics, one out of 100 NIs shows a narrow ZPL with low background, indicating a single SiV center.

\begin{figure}
\includegraphics[width=7.5cm]{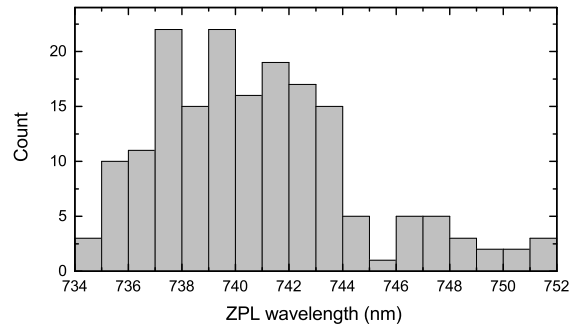}
\caption{Histogramm of the observed ZPL positions. Spectra with single and several narrow lines where taken into account.\label{FigHist} }
\end{figure}

A histogramm of the positions of the narrow ZPLs allows us to compare this sample to previously used randomly structured NI samples with regard to strain in the material which is responsible for shifting the ZPL position.\cite{Sternschulte1994,Neu2011b} Figure \ref{FigHist} shows a spread of the ZPL positions over a range of $18~$nm with a mean value of $740.9~$nm and a standard deviation of $3.9~$nm. This distribution is slightly narrower than observed for randomly structured NIs in Ref.\ \onlinecite{Neu2011b}. Furthermore, the peak of the distribution is closer to the undisturbed SiV ZPL wavelength ($738~$nm) and the number of centers with strongly shifted ZPLs ($>744~$nm) is reduced. We attribute the shift of the spectral position of the ZPL to stress of the diamond lattice at the position of the color center. A minor fraction can be explained by the biaxial macro stress of about $-0.6~$GPa resulting from thermal misfit between diamond and silicon. The major contribution stems from the micro stress of typically several GPa that can be derived from the line broadening (width of about $15~$cm$^{-1}$) in Raman measurements of thin heteroepitaxial diamond layers.\cite{Schreck2000} Its formation is characteristic for the merging of the individual grains during the early stage of film growth.  However, elastic relaxation can reduce strain for nanostructures.\cite{Schreck2000,Neu2011b}

The observed spread of ZPL positions leads to the conclusion that the sample quality is overall comparable, if not even better, to previously investigated NDs\cite{Neu2011} and NIs\cite{Neu2011b}where bright single photon emission has been demonstrated. We note that the emission rates observed here are significantly lower than previously observed (on the order of several 10000 cps). The origin of the reduced brightness is currently under investigation.

To identify single SiV centers, we measure the intensity autocorrelation function $g^{(2)}$ of the PL from selected NIs on our sample. Figure \ref{FigZPLg2}(b) shows the $g^{(2)}$ function of an emitter with a line width of 1 nm [see Fig.\ \ref{FigZPLg2}(a)]. We observe strong antibunching with $g^{(2)}_{\text{meas}}(0)=0.31$. We fit the data with the $g^{(2)}$ function of a 2 level system\cite{Neu2014b} convoluted with the instrument response of our HBT setup which is dominated by the APDs timing jitter. We find excellent agreement with the measured data (residual deviation $\Delta g^{(2)}(0)=0.1)$. Thus, the observed $g^{(2)}_{\text{meas}}(0)$ value is caused by the timing uncertainty of the APDs in conjunction with the short lifetime of the SiV center ($\tau_1=0.7~$ns) and we have proven almost pure single photon emission.

\begin{figure}[t]
\includegraphics[width=8.5cm]{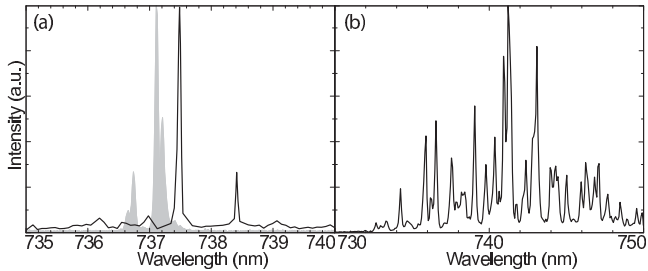}
\caption{Spectra at $10~$K showing (a) a dominant line with an underlying ensemble. The gray spectrum has been measured on a low strain CVD sample (compare Ref.\ \onlinecite{Neu2013}) (b) a large number of narrow lines\label{Figcryo} }
\end{figure}

For cryogenic temperatures ($<40~$K), the SiV ZPL exhibits a characteristic four line fine structure\cite{Sternschulte1994} in diamond samples with low crystal strain [see gray spectrum in Fig.\ \ref{Figcryo}(a)]. At $10~$K, we find spectra dominated by single spectrometer resolution-limited lines [see Fig.\ \ref{Figcryo}(a)]. These lines most probably represent fine structure lines of single SiV centers modified by crystal strain. Strain can induce a modification of the four line fine structure via changing thermal population of sublevels and thus quenching some of the lines.\cite{Sternschulte1994,Neu2013} Furthermore, we often obtain spectra consisting of many narrow lines, arising from an ensemble of SiV centers inhomogeneously broadened by crystal strain [see Fig.\ \ref{Figcryo}(b)]. Such spectra indicate a comparably high concentration of SiV centers in our sample,\cite{Neu2011a} caused by residual Si contamination of the CVD chamber. The observation of spectrometer resolution-limited lines indicates the absence of strong spectral diffusion\cite{Neu2013} due to a low influence of other impurities, which is a further sign for high diamond quality.

In summary, we demonstrate the growth of large, regular arrays of (100) oriented heteroepitaxial diamond NIs on Ir substrates.  
Individual SiV centers in our nanostructures show narrow ZPLs with a linewidth down to $1~$nm. We have thus grown arrays of nanostructures with readdressable emitters for which diamond quality is consistent with previous ND based SiV single photon sources. Further reducing the Si concentration during growth can lead to a higher yield of single centers. Our material system furthermore allows for very efficient collection of light emitted by single color centers thus facilitating their use as single photon sources or sensors. As our approach is able to create almost arbitrary (planar) structures,  we envisage the direct growth of photonic structures as e.g. waveguides\cite{Hausmann2012} on chip. This bottom up-approach to the fabrication of such nanophotonic components can ease fabrication and broaden applicability of the devices. 

This research has been partially funded by the European Commissionʼs 7. Framework Program (FP7/2007-2013) under grant agreement number 611143 (DIADEMS). EN acknowledges funding via the NanoMatFutur program of the german
ministry of education and research.
%

\clearpage

\textbf{Supplementary material for: Site selective growth of heteroepitaxial diamond nanoislands containing single SiV centers}

\textit{The supplementary material summarizes additional images of the employed etch mask as well as some additional data on SiV spectra.}

\noindent

\floatplacement{figure}{H}
\renewcommand\thefigure{S\arabic{figure}}
\setcounter{figure}{0}
\textbf{Additional data on SiV spectra:}

\begin{figure}[H]
\includegraphics[width=8.5cm]{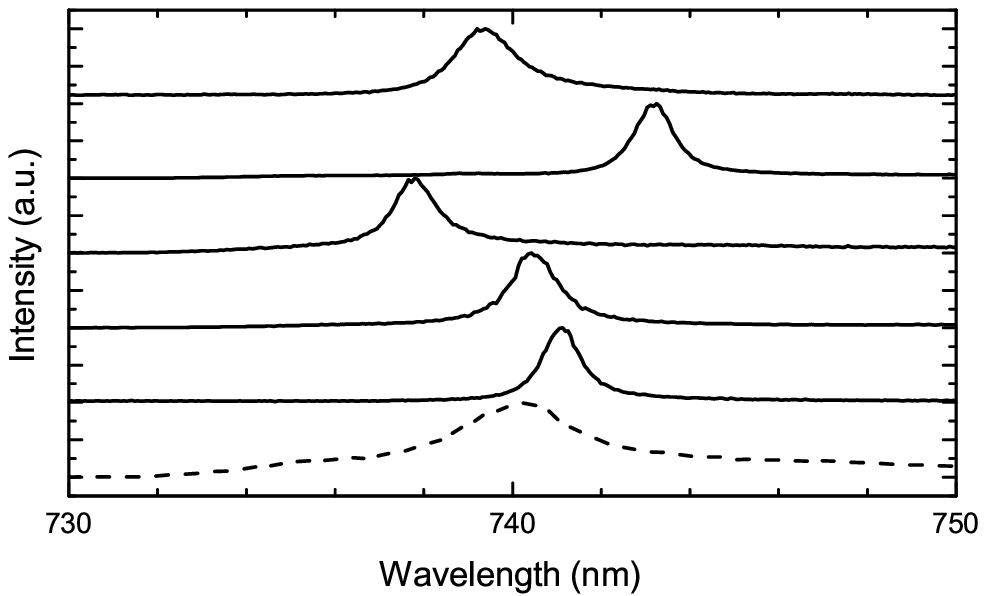}
\caption{ZPLs of several SiV color centers at room temperature. The spectra have been normalized and stacked for clarity. The dashed line shows an ensemble spectrum with a linewidth of $4~$nm.\label{FigZPLstacked} }
\end{figure}

Figure \ref{FigZPLstacked} provides in addition to Figs.\ 4 and 5 in the manuscript information on the spectra of SiV centers in our nanoislands. Note the strongly dominating zero phonon line (ZPL) with almost no sideband emission for all spectra observed. The dashed line shows the ZPL emission of an ensemble of SiV centers in our samples, which is significantly broader than the single emitter lines. 
   
\textbf{Additional information on structuring process:}

Figure \ref{Figetchedmask1} shows an example of an HSQ mask after $60~$s of etching. No discernible mask erosion has occurred during the etch. Similar HSQ masks have been used to etch diamond nanowires and are know to withstand the employed plasma for more than $10~$mins.\cite{Neu2014} The ICP-RIE etch leads to a slight re-deposition of Ir on the sidewall of the HSQ mask. This deposit is not detrimental for the process. Only after some minutes of etching, it can lead to the formation of Ir flakes after mask removal. Our growth experiments show that these flakes are not disturbing heteroepitaxial diamond growth. Fig.\ \ref{Figetchedmask1} shows a mask employed to pattern a nanoisland, also here no mask erosion is discernible. In the future, the process can be optimized by using much thinner etch masks (allowing typically for an enhanced resist adhesion) as well as minimized etch times. 

\begin{figure}[H]
\includegraphics[width=8.5cm]{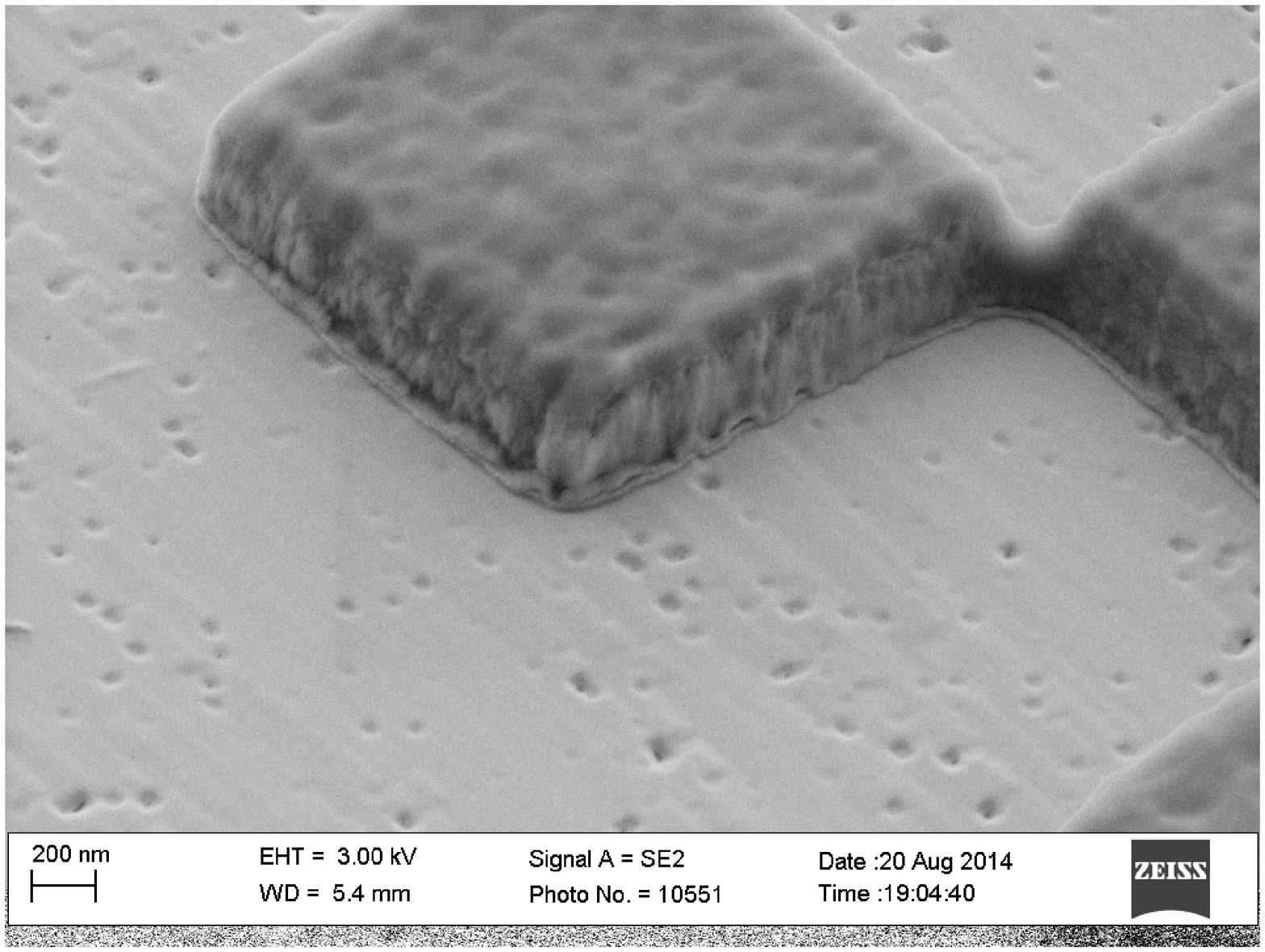}
\caption{HSQ mask of a larger structure (chess board) etched for $60~$s. No significant mask erosion is visible. \label{Figetchedmask1} }
\end{figure}
\begin{figure}[H]
\includegraphics[width=8.5cm]{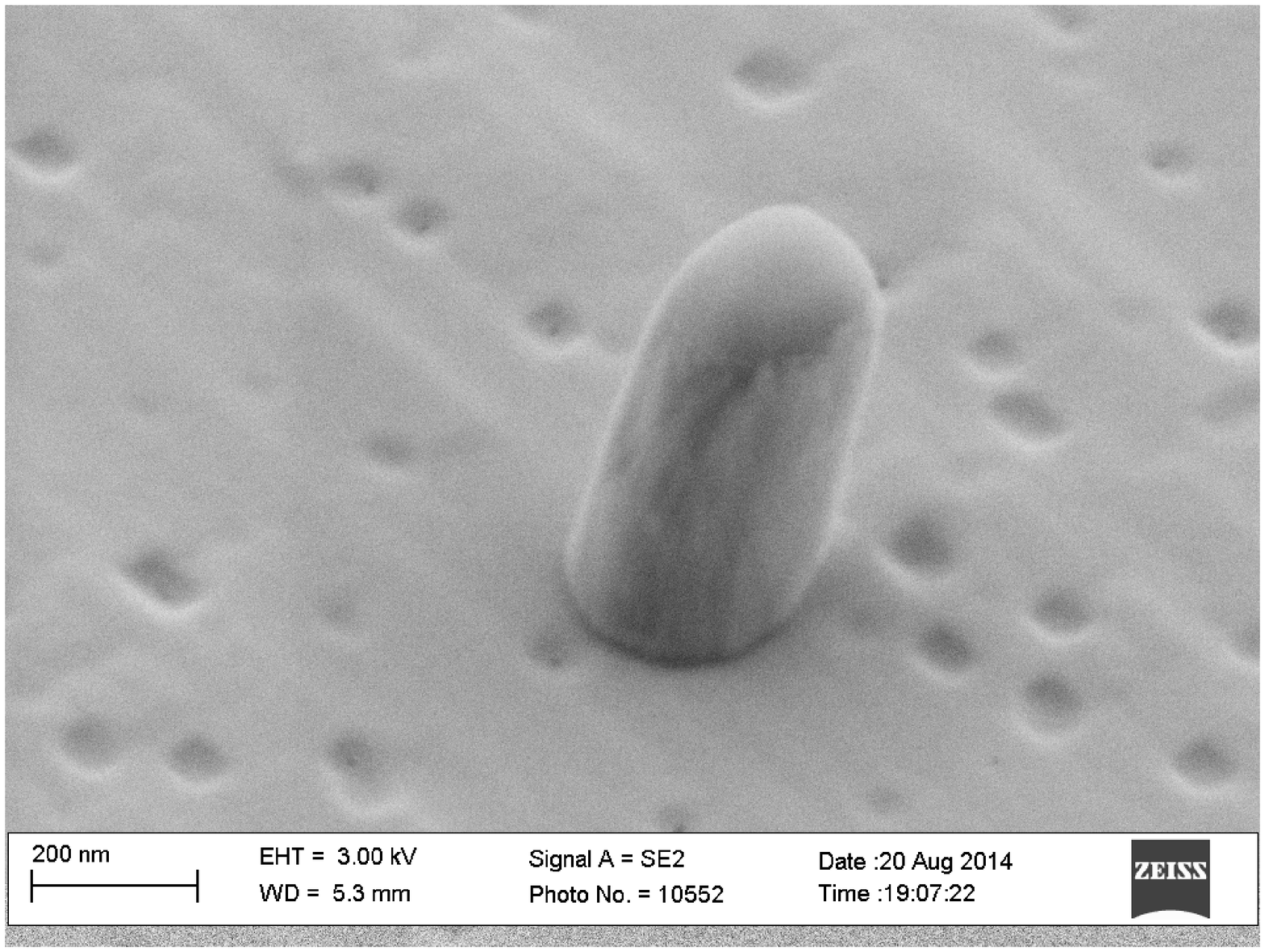}
\caption{HSQ mask used to pattern a nanoisland after a $60~$s etch. No significant mask erosion is visible. \label{Figetchedmask1} }
\end{figure}


\begin{thebibliography}{26}%
\makeatletter
\providecommand \@ifxundefined [1]{%
 \@ifx{#1\undefined}
}%
\providecommand \@ifnum [1]{%
 \ifnum #1\expandafter \@firstoftwo
 \else \expandafter \@secondoftwo
 \fi
}%
\providecommand \@ifx [1]{%
 \ifx #1\expandafter \@firstoftwo
 \else \expandafter \@secondoftwo
 \fi
}%
\providecommand \natexlab [1]{#1}%
\providecommand \enquote  [1]{``#1''}%
\providecommand \bibnamefont  [1]{#1}%
\providecommand \bibfnamefont [1]{#1}%
\providecommand \citenamefont [1]{#1}%
\providecommand \href@noop [0]{\@secondoftwo}%
\providecommand \href [0]{\begingroup \@sanitize@url \@href}%
\providecommand \@href[1]{\@@startlink{#1}\@@href}%
\providecommand \@@href[1]{\endgroup#1\@@endlink}%
\providecommand \@sanitize@url [0]{\catcode `\\12\catcode `\$12\catcode
  `\&12\catcode `\#12\catcode `\^12\catcode `\_12\catcode `\%12\relax}%
\providecommand \@@startlink[1]{}%
\providecommand \@@endlink[0]{}%
\providecommand \url  [0]{\begingroup\@sanitize@url \@url }%
\providecommand \@url [1]{\endgroup\@href {#1}{\urlprefix }}%
\providecommand \urlprefix  [0]{URL }%
\providecommand \Eprint [0]{\href }%
\providecommand \doibase [0]{http://dx.doi.org/}%
\providecommand \selectlanguage [0]{\@gobble}%
\providecommand \bibinfo  [0]{\@secondoftwo}%
\providecommand \bibfield  [0]{\@secondoftwo}%
\providecommand \translation [1]{[#1]}%
\providecommand \BibitemOpen [0]{}%
\providecommand \bibitemStop [0]{}%
\providecommand \bibitemNoStop [0]{.\EOS\space}%
\providecommand \EOS [0]{\spacefactor3000\relax}%
\providecommand \BibitemShut  [1]{\csname bibitem#1\endcsname}%
\let\auto@bib@innerbib\@empty
\bibitem [{\citenamefont {Neu}\ and\ \citenamefont {Becher}(2014)}]{Neu2014b}%
  \BibitemOpen
  \bibfield  {author} {\bibinfo {author} {\bibfnamefont {E.}~\bibnamefont
  {Neu}}\ and\ \bibinfo {author} {\bibfnamefont {C.}~\bibnamefont {Becher}},\
  }\enquote {\bibinfo {title} {Diamond- based single- photon sources and their
  application in quantum key distribution},}\ \ (\bibinfo  {publisher}
  {Woodhead Publishing, Cambridge},\ \bibinfo {year} {2014})\ Chap.~\bibinfo
  {chapter} {6}, pp.\ \bibinfo {pages} {127--159}\BibitemShut {NoStop}%
\bibitem [{\citenamefont {Rondin}\ \emph {et~al.}(2014)\citenamefont {Rondin},
  \citenamefont {Tetienne}, \citenamefont {Hingant}, \citenamefont {Roch},
  \citenamefont {Maletinsky},\ and\ \citenamefont {Jacques}}]{Rondin2014}%
  \BibitemOpen
  \bibfield  {author} {\bibinfo {author} {\bibfnamefont {L.}~\bibnamefont
  {Rondin}}, \bibinfo {author} {\bibfnamefont {J.-P.}\ \bibnamefont
  {Tetienne}}, \bibinfo {author} {\bibfnamefont {T.}~\bibnamefont {Hingant}},
  \bibinfo {author} {\bibfnamefont {J.-F.}\ \bibnamefont {Roch}}, \bibinfo
  {author} {\bibfnamefont {P.}~\bibnamefont {Maletinsky}}, \ and\ \bibinfo
  {author} {\bibfnamefont {V.}~\bibnamefont {Jacques}},\ }\href
  {http://stacks.iop.org/0034-4885/77/i=5/a=056503} {\bibfield  {journal}
  {\bibinfo  {journal} {Reports on Progress in Physics}\ }\textbf {\bibinfo
  {volume} {77}},\ \bibinfo {pages} {056503} (\bibinfo {year}
  {2014})}\BibitemShut {NoStop}%
\bibitem [{\citenamefont {Tisler}\ \emph {et~al.}(2013)\citenamefont {Tisler},
  \citenamefont {Oeckinghaus}, \citenamefont {St\"ohr}, \citenamefont
  {Kolesov}, \citenamefont {Reuter}, \citenamefont {Reinhard},\ and\
  \citenamefont {Wrachtrup}}]{Tisler2013a}%
  \BibitemOpen
  \bibfield  {author} {\bibinfo {author} {\bibfnamefont {J.}~\bibnamefont
  {Tisler}}, \bibinfo {author} {\bibfnamefont {T.}~\bibnamefont {Oeckinghaus}},
  \bibinfo {author} {\bibfnamefont {R.~J.}\ \bibnamefont {St\"ohr}}, \bibinfo
  {author} {\bibfnamefont {R.}~\bibnamefont {Kolesov}}, \bibinfo {author}
  {\bibfnamefont {R.}~\bibnamefont {Reuter}}, \bibinfo {author} {\bibfnamefont
  {F.}~\bibnamefont {Reinhard}}, \ and\ \bibinfo {author} {\bibfnamefont
  {J.}~\bibnamefont {Wrachtrup}},\ }\href {\doibase 10.1021/nl401129m}
  {\bibfield  {journal} {\bibinfo  {journal} {Nano Letters}\ }\textbf {\bibinfo
  {volume} {13}},\ \bibinfo {pages} {3152} (\bibinfo {year}
  {2013})}\BibitemShut {NoStop}%
\bibitem [{\citenamefont {Aharonovich}\ and\ \citenamefont
  {Neu}(2014)}]{Aharonovich2014a}%
  \BibitemOpen
  \bibfield  {author} {\bibinfo {author} {\bibfnamefont {I.}~\bibnamefont
  {Aharonovich}}\ and\ \bibinfo {author} {\bibfnamefont {E.}~\bibnamefont
  {Neu}},\ }\href {\doibase 10.1002/adom.201400189} {\bibfield  {journal}
  {\bibinfo  {journal} {Advanced Optical Materials}\ }\textbf {\bibinfo
  {volume} {2}},\ \bibinfo {pages} {911–928} (\bibinfo {year}
  {2014})}\BibitemShut {NoStop}%
\bibitem [{\citenamefont {Babinec}\ \emph {et~al.}(2010)\citenamefont
  {Babinec}, \citenamefont {Hausmann}, \citenamefont {Khan}, \citenamefont
  {Zhang}, \citenamefont {Maze}, \citenamefont {Hemmer},\ and\ \citenamefont
  {Loncar}}]{Babinec2010}%
  \BibitemOpen
  \bibfield  {author} {\bibinfo {author} {\bibfnamefont {T.}~\bibnamefont
  {Babinec}}, \bibinfo {author} {\bibfnamefont {B.}~\bibnamefont {Hausmann}},
  \bibinfo {author} {\bibfnamefont {M.}~\bibnamefont {Khan}}, \bibinfo {author}
  {\bibfnamefont {Y.}~\bibnamefont {Zhang}}, \bibinfo {author} {\bibfnamefont
  {J.}~\bibnamefont {Maze}}, \bibinfo {author} {\bibfnamefont {P.}~\bibnamefont
  {Hemmer}}, \ and\ \bibinfo {author} {\bibfnamefont {M.}~\bibnamefont
  {Loncar}},\ }\href {\doibase 10.1038/NNANO.2010.6} {\bibfield  {journal}
  {\bibinfo  {journal} {Nature Nanotech.}\ }\textbf {\bibinfo {volume} {5}},\
  \bibinfo {pages} {195} (\bibinfo {year} {2010})}\BibitemShut {NoStop}%
\bibitem [{\citenamefont {Aharonovich}\ \emph {et~al.}(2013)\citenamefont
  {Aharonovich}, \citenamefont {Lee}, \citenamefont {Magyar}, \citenamefont
  {Bracher},\ and\ \citenamefont {Hu}}]{Aharonovich2013}%
  \BibitemOpen
  \bibfield  {author} {\bibinfo {author} {\bibfnamefont {I.}~\bibnamefont
  {Aharonovich}}, \bibinfo {author} {\bibfnamefont {J.~C.}\ \bibnamefont
  {Lee}}, \bibinfo {author} {\bibfnamefont {A.~P.}\ \bibnamefont {Magyar}},
  \bibinfo {author} {\bibfnamefont {D.~O.}\ \bibnamefont {Bracher}}, \ and\
  \bibinfo {author} {\bibfnamefont {E.~L.}\ \bibnamefont {Hu}},\ }\href
  {\doibase 10.1002/lpor.201300065} {\bibfield  {journal} {\bibinfo  {journal}
  {Laser \& Photonics Reviews}\ }\textbf {\bibinfo {volume} {7}},\ \bibinfo
  {pages} {L61} (\bibinfo {year} {2013})}\BibitemShut {NoStop}%
\bibitem [{\citenamefont {Sovyk}\ \emph {et~al.}(2014)\citenamefont {Sovyk},
  \citenamefont {Ralchenko}, \citenamefont {Komlenok}, \citenamefont {Khomich},
  \citenamefont {Shershulin}, \citenamefont {Vorobyov}, \citenamefont {Vlasov},
  \citenamefont {Konov},\ and\ \citenamefont {Akimov}}]{Sovyk2014}%
  \BibitemOpen
  \bibfield  {author} {\bibinfo {author} {\bibfnamefont {D.}~\bibnamefont
  {Sovyk}}, \bibinfo {author} {\bibfnamefont {V.}~\bibnamefont {Ralchenko}},
  \bibinfo {author} {\bibfnamefont {M.}~\bibnamefont {Komlenok}}, \bibinfo
  {author} {\bibfnamefont {A.}~\bibnamefont {Khomich}}, \bibinfo {author}
  {\bibfnamefont {V.}~\bibnamefont {Shershulin}}, \bibinfo {author}
  {\bibfnamefont {V.}~\bibnamefont {Vorobyov}}, \bibinfo {author}
  {\bibfnamefont {I.}~\bibnamefont {Vlasov}}, \bibinfo {author} {\bibfnamefont
  {V.}~\bibnamefont {Konov}}, \ and\ \bibinfo {author} {\bibfnamefont
  {A.}~\bibnamefont {Akimov}},\ }\href {\doibase 10.1007/s00339-014-8877-2}
  {\bibfield  {journal} {\bibinfo  {journal} {Appl. Phys. A}\ } (\bibinfo
  {year} {2014}),\ 10.1007/s00339-014-8877-2}\BibitemShut {NoStop}%
\bibitem [{\citenamefont {Masuda}\ \emph {et~al.}(2001)\citenamefont {Masuda},
  \citenamefont {Yanagishita}, \citenamefont {Yasui}, \citenamefont {Nishio},
  \citenamefont {Yagi}, \citenamefont {Rao},\ and\ \citenamefont
  {Fujishima}}]{Masuda2001}%
  \BibitemOpen
  \bibfield  {author} {\bibinfo {author} {\bibfnamefont {H.}~\bibnamefont
  {Masuda}}, \bibinfo {author} {\bibfnamefont {T.}~\bibnamefont {Yanagishita}},
  \bibinfo {author} {\bibfnamefont {K.}~\bibnamefont {Yasui}}, \bibinfo
  {author} {\bibfnamefont {K.}~\bibnamefont {Nishio}}, \bibinfo {author}
  {\bibfnamefont {I.}~\bibnamefont {Yagi}}, \bibinfo {author} {\bibfnamefont
  {T.~N.}\ \bibnamefont {Rao}}, \ and\ \bibinfo {author} {\bibfnamefont
  {A.}~\bibnamefont {Fujishima}},\ }\href {\doibase
  10.1002/1521-4095(200102)13:4<247::AID-ADMA247>3.0.CO;2-H} {\bibfield
  {journal} {\bibinfo  {journal} {Advanced Materials}\ }\textbf {\bibinfo
  {volume} {13}},\ \bibinfo {pages} {247} (\bibinfo {year} {2001})}\BibitemShut
  {NoStop}%
\bibitem [{\citenamefont {Furuyama}\ \emph {et~al.}(2015)\citenamefont
  {Furuyama}, \citenamefont {Tahara}, \citenamefont {Iwasaki}, \citenamefont
  {Shimizu}, \citenamefont {Yaita}, \citenamefont {Kondo}, \citenamefont
  {Kodera},\ and\ \citenamefont {Hatano}}]{Furuyama2015}%
  \BibitemOpen
  \bibfield  {author} {\bibinfo {author} {\bibfnamefont {S.}~\bibnamefont
  {Furuyama}}, \bibinfo {author} {\bibfnamefont {K.}~\bibnamefont {Tahara}},
  \bibinfo {author} {\bibfnamefont {T.}~\bibnamefont {Iwasaki}}, \bibinfo
  {author} {\bibfnamefont {M.}~\bibnamefont {Shimizu}}, \bibinfo {author}
  {\bibfnamefont {J.}~\bibnamefont {Yaita}}, \bibinfo {author} {\bibfnamefont
  {M.}~\bibnamefont {Kondo}}, \bibinfo {author} {\bibfnamefont
  {T.}~\bibnamefont {Kodera}}, \ and\ \bibinfo {author} {\bibfnamefont
  {M.}~\bibnamefont {Hatano}},\ }\href {\doibase 10.1063/1.4933103} {\bibfield
  {journal} {\bibinfo  {journal} {Applied Physics Letters}\ }\textbf {\bibinfo
  {volume} {107}},\ \bibinfo {eid} {163102} (\bibinfo {year}
  {2015})}\BibitemShut {NoStop}%
\bibitem [{\citenamefont {Shimoni}\ \emph {et~al.}(2014)\citenamefont
  {Shimoni}, \citenamefont {Cervenka}, \citenamefont {Karle}, \citenamefont
  {Fox}, \citenamefont {Gibson}, \citenamefont {Tomljenovic-Hanic},
  \citenamefont {Greentree},\ and\ \citenamefont {Prawer}}]{Shimoni2014}%
  \BibitemOpen
  \bibfield  {author} {\bibinfo {author} {\bibfnamefont {O.}~\bibnamefont
  {Shimoni}}, \bibinfo {author} {\bibfnamefont {J.}~\bibnamefont {Cervenka}},
  \bibinfo {author} {\bibfnamefont {T.~J.}\ \bibnamefont {Karle}}, \bibinfo
  {author} {\bibfnamefont {K.}~\bibnamefont {Fox}}, \bibinfo {author}
  {\bibfnamefont {B.~C.}\ \bibnamefont {Gibson}}, \bibinfo {author}
  {\bibfnamefont {S.}~\bibnamefont {Tomljenovic-Hanic}}, \bibinfo {author}
  {\bibfnamefont {A.~D.}\ \bibnamefont {Greentree}}, \ and\ \bibinfo {author}
  {\bibfnamefont {S.}~\bibnamefont {Prawer}},\ }\href {\doibase
  10.1021/am5016556} {\bibfield  {journal} {\bibinfo  {journal} {ACS Applied
  Materials \& Interfaces}\ }\textbf {\bibinfo {volume} {6}},\ \bibinfo {pages}
  {8894} (\bibinfo {year} {2014})}\BibitemShut {NoStop}%
\bibitem [{\citenamefont {Singh}\ \emph {et~al.}(2014)\citenamefont {Singh},
  \citenamefont {Thomas}, \citenamefont {Martyshkin}, \citenamefont
  {Kozlovskaya}, \citenamefont {Kharlampieva},\ and\ \citenamefont
  {Catledge}}]{Singh2014}%
  \BibitemOpen
  \bibfield  {author} {\bibinfo {author} {\bibfnamefont {S.}~\bibnamefont
  {Singh}}, \bibinfo {author} {\bibfnamefont {V.}~\bibnamefont {Thomas}},
  \bibinfo {author} {\bibfnamefont {D.}~\bibnamefont {Martyshkin}}, \bibinfo
  {author} {\bibfnamefont {V.}~\bibnamefont {Kozlovskaya}}, \bibinfo {author}
  {\bibfnamefont {E.}~\bibnamefont {Kharlampieva}}, \ and\ \bibinfo {author}
  {\bibfnamefont {S.~A.}\ \bibnamefont {Catledge}},\ }\href
  {http://stacks.iop.org/0957-4484/25/i=4/a=045302} {\bibfield  {journal}
  {\bibinfo  {journal} {Nanotechnology}\ }\textbf {\bibinfo {volume} {25}},\
  \bibinfo {pages} {045302} (\bibinfo {year} {2014})}\BibitemShut {NoStop}%
\bibitem [{\citenamefont {Gsell}\ \emph {et~al.}(2004)\citenamefont {Gsell},
  \citenamefont {Bauer}, \citenamefont {Goldfu{\ss}}, \citenamefont {Schreck},\
  and\ \citenamefont {Stritzker}}]{Gsell2004}%
  \BibitemOpen
  \bibfield  {author} {\bibinfo {author} {\bibfnamefont {S.}~\bibnamefont
  {Gsell}}, \bibinfo {author} {\bibfnamefont {T.}~\bibnamefont {Bauer}},
  \bibinfo {author} {\bibfnamefont {J.}~\bibnamefont {Goldfu{\ss}}}, \bibinfo
  {author} {\bibfnamefont {M.}~\bibnamefont {Schreck}}, \ and\ \bibinfo
  {author} {\bibfnamefont {B.}~\bibnamefont {Stritzker}},\ }\href {\doibase
  10.1063/1.1758780} {\bibfield  {journal} {\bibinfo  {journal} {Appl. Phys.
  Lett.}\ }\textbf {\bibinfo {volume} {84}},\ \bibinfo {pages} {4541} (\bibinfo
  {year} {2004})}\BibitemShut {NoStop}%
\bibitem [{\citenamefont {Ando}\ \emph {et~al.}(2004)\citenamefont {Ando},
  \citenamefont {Kuwabara}, \citenamefont {Suzuki},\ and\ \citenamefont
  {Sawabe}}]{Ando2004}%
  \BibitemOpen
  \bibfield  {author} {\bibinfo {author} {\bibfnamefont {Y.}~\bibnamefont
  {Ando}}, \bibinfo {author} {\bibfnamefont {J.}~\bibnamefont {Kuwabara}},
  \bibinfo {author} {\bibfnamefont {K.}~\bibnamefont {Suzuki}}, \ and\ \bibinfo
  {author} {\bibfnamefont {A.}~\bibnamefont {Sawabe}},\ }\href {\doibase
  10.1016/j.diamond.2004.06.025} {\bibfield  {journal} {\bibinfo  {journal}
  {Diam. Relat. Mater.}\ }\textbf {\bibinfo {volume} {{13}}},\ \bibinfo {pages}
  {1975} (\bibinfo {year} {{2004}})}\BibitemShut {NoStop}%
\bibitem [{\citenamefont {Ando}\ \emph {et~al.}(2012)\citenamefont {Ando},
  \citenamefont {Kamano}, \citenamefont {Suzuki},\ and\ \citenamefont
  {Sawabe}}]{Ando2012}%
  \BibitemOpen
  \bibfield  {author} {\bibinfo {author} {\bibfnamefont {Y.}~\bibnamefont
  {Ando}}, \bibinfo {author} {\bibfnamefont {T.}~\bibnamefont {Kamano}},
  \bibinfo {author} {\bibfnamefont {K.}~\bibnamefont {Suzuki}}, \ and\ \bibinfo
  {author} {\bibfnamefont {A.}~\bibnamefont {Sawabe}},\ }\href {\doibase
  10.1143/JJAP.51.090101} {\bibfield  {journal} {\bibinfo  {journal} {Japanese
  Journal of Applied Physics}\ }\textbf {\bibinfo {volume} {51}},\ \bibinfo
  {pages} {090101} (\bibinfo {year} {2012})}\BibitemShut {NoStop}%
\bibitem [{\citenamefont {Neu}\ \emph {et~al.}(2011{\natexlab{a}})\citenamefont
  {Neu}, \citenamefont {Steinmetz}, \citenamefont {Riedrich-M\"oller},
  \citenamefont {Gsell}, \citenamefont {Fischer}, \citenamefont {Schreck},\
  and\ \citenamefont {Becher}}]{Neu2011}%
  \BibitemOpen
  \bibfield  {author} {\bibinfo {author} {\bibfnamefont {E.}~\bibnamefont
  {Neu}}, \bibinfo {author} {\bibfnamefont {D.}~\bibnamefont {Steinmetz}},
  \bibinfo {author} {\bibfnamefont {J.}~\bibnamefont {Riedrich-M\"oller}},
  \bibinfo {author} {\bibfnamefont {S.}~\bibnamefont {Gsell}}, \bibinfo
  {author} {\bibfnamefont {M.}~\bibnamefont {Fischer}}, \bibinfo {author}
  {\bibfnamefont {M.}~\bibnamefont {Schreck}}, \ and\ \bibinfo {author}
  {\bibfnamefont {C.}~\bibnamefont {Becher}},\ }\href {\doibase
  10.1088/1367-2630/13/2/025012} {\bibfield  {journal} {\bibinfo  {journal}
  {New J. Phys.}\ }\textbf {\bibinfo {volume} {13}},\ \bibinfo {pages} {025012}
  (\bibinfo {year} {2011}{\natexlab{a}})}\BibitemShut {NoStop}%
\bibitem [{\citenamefont {Neu}\ \emph {et~al.}(2012)\citenamefont {Neu},
  \citenamefont {Agio},\ and\ \citenamefont {Becher}}]{Neu2012a}%
  \BibitemOpen
  \bibfield  {author} {\bibinfo {author} {\bibfnamefont {E.}~\bibnamefont
  {Neu}}, \bibinfo {author} {\bibfnamefont {M.}~\bibnamefont {Agio}}, \ and\
  \bibinfo {author} {\bibfnamefont {C.}~\bibnamefont {Becher}},\ }\href
  {\doibase 10.1364/OE.20.019956} {\bibfield  {journal} {\bibinfo  {journal}
  {Opt. Express}\ }\textbf {\bibinfo {volume} {20}},\ \bibinfo {pages} {19956}
  (\bibinfo {year} {2012})}\BibitemShut {NoStop}%
\bibitem [{\citenamefont {Neu}\ \emph {et~al.}(2011{\natexlab{b}})\citenamefont
  {Neu}, \citenamefont {Fischer}, \citenamefont {Gsell}, \citenamefont
  {Schreck},\ and\ \citenamefont {Becher}}]{Neu2011b}%
  \BibitemOpen
  \bibfield  {author} {\bibinfo {author} {\bibfnamefont {E.}~\bibnamefont
  {Neu}}, \bibinfo {author} {\bibfnamefont {M.}~\bibnamefont {Fischer}},
  \bibinfo {author} {\bibfnamefont {S.}~\bibnamefont {Gsell}}, \bibinfo
  {author} {\bibfnamefont {M.}~\bibnamefont {Schreck}}, \ and\ \bibinfo
  {author} {\bibfnamefont {C.}~\bibnamefont {Becher}},\ }\href {\doibase
  10.1103/PhysRevB.84.205211} {\bibfield  {journal} {\bibinfo  {journal} {Phys.
  Rev. B}\ }\textbf {\bibinfo {volume} {84}},\ \bibinfo {pages} {205211}
  (\bibinfo {year} {2011}{\natexlab{b}})}\BibitemShut {NoStop}%
\bibitem [{\citenamefont {Fischer}\ \emph {et~al.}(2008)\citenamefont
  {Fischer}, \citenamefont {Gsell}, \citenamefont {Schreck}, \citenamefont
  {Brescia},\ and\ \citenamefont {Stritzker}}]{Fischer2008a}%
  \BibitemOpen
  \bibfield  {author} {\bibinfo {author} {\bibfnamefont {M.}~\bibnamefont
  {Fischer}}, \bibinfo {author} {\bibfnamefont {S.}~\bibnamefont {Gsell}},
  \bibinfo {author} {\bibfnamefont {M.}~\bibnamefont {Schreck}}, \bibinfo
  {author} {\bibfnamefont {R.}~\bibnamefont {Brescia}}, \ and\ \bibinfo
  {author} {\bibfnamefont {B.}~\bibnamefont {Stritzker}},\ }\href {\doibase
  10.1016/j.diamond.2008.02.028} {\bibfield  {journal} {\bibinfo  {journal}
  {Diam. Relat. Mater.}\ }\textbf {\bibinfo {volume} {17}},\ \bibinfo {pages}
  {1035} (\bibinfo {year} {2008})}\BibitemShut {NoStop}%
\bibitem [{\citenamefont {Brescia}\ \emph {et~al.}(2008)\citenamefont
  {Brescia}, \citenamefont {Schreck}, \citenamefont {Gsell},\ and\
  \citenamefont {Stritzker}}]{Brescia2008}%
  \BibitemOpen
  \bibfield  {author} {\bibinfo {author} {\bibfnamefont {R.}~\bibnamefont
  {Brescia}}, \bibinfo {author} {\bibfnamefont {M.}~\bibnamefont {Schreck}},
  \bibinfo {author} {\bibfnamefont {M.}~\bibnamefont {Gsell}, \bibfnamefont
  {S.and~Fischer}}, \ and\ \bibinfo {author} {\bibfnamefont {B.}~\bibnamefont
  {Stritzker}},\ }\href
  {http://www.scopus.com/inward/record.url?eid=2-s2.0-48849104213&partnerID=40&md5=6965a111ae5603c4851478d7417f68d1}
  {\bibfield  {journal} {\bibinfo  {journal} {Diam. Relat. Mater.}\ }\textbf
  {\bibinfo {volume} {17}},\ \bibinfo {pages} {1045} (\bibinfo {year}
  {2008})}\BibitemShut {NoStop}%
\bibitem [{sup()}]{suppl2015}%
  \BibitemOpen
  \href@noop {} { {\bibinfo {title} {See supplementary material below for additional spectra of SiV centers and images of HSQ
  mask after ICP-RIE etch}}\ }\BibitemShut {NoStop}%
\bibitem [{\citenamefont {Sternschulte}\ \emph {et~al.}(1994)\citenamefont
  {Sternschulte}, \citenamefont {Thonke}, \citenamefont {Sauer}, \citenamefont
  {M\"unzinger},\ and\ \citenamefont {Michler}}]{Sternschulte1994}%
  \BibitemOpen
  \bibfield  {author} {\bibinfo {author} {\bibfnamefont {H.}~\bibnamefont
  {Sternschulte}}, \bibinfo {author} {\bibfnamefont {K.}~\bibnamefont
  {Thonke}}, \bibinfo {author} {\bibfnamefont {R.}~\bibnamefont {Sauer}},
  \bibinfo {author} {\bibfnamefont {P.~C.}\ \bibnamefont {M\"unzinger}}, \ and\
  \bibinfo {author} {\bibfnamefont {P.}~\bibnamefont {Michler}},\ }\href
  {\doibase 10.1103/PhysRevB.50.14554} {\bibfield  {journal} {\bibinfo
  {journal} {Phys. Rev. B}\ }\textbf {\bibinfo {volume} {50}},\ \bibinfo
  {pages} {14554} (\bibinfo {year} {1994})}\BibitemShut {NoStop}%
\bibitem [{\citenamefont {Schreck}\ \emph {et~al.}(2000)\citenamefont
  {Schreck}, \citenamefont {Roll}, \citenamefont {Michler}, \citenamefont
  {Blank},\ and\ \citenamefont {Stritzker}}]{Schreck2000}%
  \BibitemOpen
  \bibfield  {author} {\bibinfo {author} {\bibfnamefont {M.}~\bibnamefont
  {Schreck}}, \bibinfo {author} {\bibfnamefont {H.}~\bibnamefont {Roll}},
  \bibinfo {author} {\bibfnamefont {J.}~\bibnamefont {Michler}}, \bibinfo
  {author} {\bibfnamefont {E.}~\bibnamefont {Blank}}, \ and\ \bibinfo {author}
  {\bibfnamefont {B.}~\bibnamefont {Stritzker}},\ }\href@noop {} {\bibfield
  {journal} {\bibinfo  {journal} {J. Appl. Phys.}\ }\textbf {\bibinfo {volume}
  {88}},\ \bibinfo {pages} {2456} (\bibinfo {year} {2000})}\BibitemShut
  {NoStop}%
\bibitem [{\citenamefont {Neu}\ \emph {et~al.}(2013)\citenamefont {Neu},
  \citenamefont {Hepp}, \citenamefont {Hauschild}, \citenamefont {Gsell},
  \citenamefont {Fischer}, \citenamefont {Sternschulte}, \citenamefont
  {Steinmüller-Nethl}, \citenamefont {Schreck},\ and\ \citenamefont
  {Becher}}]{Neu2013}%
  \BibitemOpen
  \bibfield  {author} {\bibinfo {author} {\bibfnamefont {E.}~\bibnamefont
  {Neu}}, \bibinfo {author} {\bibfnamefont {C.}~\bibnamefont {Hepp}}, \bibinfo
  {author} {\bibfnamefont {M.}~\bibnamefont {Hauschild}}, \bibinfo {author}
  {\bibfnamefont {S.}~\bibnamefont {Gsell}}, \bibinfo {author} {\bibfnamefont
  {M.}~\bibnamefont {Fischer}}, \bibinfo {author} {\bibfnamefont
  {H.}~\bibnamefont {Sternschulte}}, \bibinfo {author} {\bibfnamefont
  {D.}~\bibnamefont {Steinmüller-Nethl}}, \bibinfo {author} {\bibfnamefont
  {M.}~\bibnamefont {Schreck}}, \ and\ \bibinfo {author} {\bibfnamefont
  {C.}~\bibnamefont {Becher}},\ }\href {\doibase 10.1088/1367-2630/15/4/043005}
  {\bibfield  {journal} {\bibinfo  {journal} {New Journal of Physics}\ }\textbf
  {\bibinfo {volume} {15}},\ \bibinfo {pages} {043005} (\bibinfo {year}
  {2013})}\BibitemShut {NoStop}%
\bibitem [{\citenamefont {Neu}\ \emph {et~al.}(2011{\natexlab{c}})\citenamefont
  {Neu}, \citenamefont {Arend}, \citenamefont {Gross}, \citenamefont {Guldner},
  \citenamefont {Hepp}, \citenamefont {Steinmetz}, \citenamefont {Zscherpel},
  \citenamefont {Ghodbane}, \citenamefont {Sternschulte}, \citenamefont
  {Steinm\"uller-Nethl}, \citenamefont {Liang}, \citenamefont {Krueger},\ and\
  \citenamefont {Becher}}]{Neu2011a}%
  \BibitemOpen
  \bibfield  {author} {\bibinfo {author} {\bibfnamefont {E.}~\bibnamefont
  {Neu}}, \bibinfo {author} {\bibfnamefont {C.}~\bibnamefont {Arend}}, \bibinfo
  {author} {\bibfnamefont {E.}~\bibnamefont {Gross}}, \bibinfo {author}
  {\bibfnamefont {F.}~\bibnamefont {Guldner}}, \bibinfo {author} {\bibfnamefont
  {C.}~\bibnamefont {Hepp}}, \bibinfo {author} {\bibfnamefont {D.}~\bibnamefont
  {Steinmetz}}, \bibinfo {author} {\bibfnamefont {E.}~\bibnamefont
  {Zscherpel}}, \bibinfo {author} {\bibfnamefont {S.}~\bibnamefont {Ghodbane}},
  \bibinfo {author} {\bibfnamefont {H.}~\bibnamefont {Sternschulte}}, \bibinfo
  {author} {\bibfnamefont {D.}~\bibnamefont {Steinm\"uller-Nethl}}, \bibinfo
  {author} {\bibfnamefont {Y.}~\bibnamefont {Liang}}, \bibinfo {author}
  {\bibfnamefont {A.}~\bibnamefont {Krueger}}, \ and\ \bibinfo {author}
  {\bibfnamefont {C.}~\bibnamefont {Becher}},\ }\href {\doibase
  10.1063/1.3599608} {\bibfield  {journal} {\bibinfo  {journal} {Appl. Phys.
  Lett.}\ }\textbf {\bibinfo {volume} {98}},\ \bibinfo {eid} {243107} (\bibinfo
  {year} {2011}{\natexlab{c}})}\BibitemShut {NoStop}%
\bibitem [{\citenamefont {Hausmann}\ \emph {et~al.}(2012)\citenamefont
  {Hausmann}, \citenamefont {Shields}, \citenamefont {Quan}, \citenamefont
  {Maletinsky}, \citenamefont {McCutcheon}, \citenamefont {Choy}, \citenamefont
  {Babinec}, \citenamefont {Kubanek}, \citenamefont {Yacoby}, \citenamefont
  {Lukin},\ and\ \citenamefont {Loncar}}]{Hausmann2012}%
  \BibitemOpen
  \bibfield  {author} {\bibinfo {author} {\bibfnamefont {B.~J.~M.}\
  \bibnamefont {Hausmann}}, \bibinfo {author} {\bibfnamefont {B.}~\bibnamefont
  {Shields}}, \bibinfo {author} {\bibfnamefont {Q.}~\bibnamefont {Quan}},
  \bibinfo {author} {\bibfnamefont {P.}~\bibnamefont {Maletinsky}}, \bibinfo
  {author} {\bibfnamefont {M.}~\bibnamefont {McCutcheon}}, \bibinfo {author}
  {\bibfnamefont {J.~T.}\ \bibnamefont {Choy}}, \bibinfo {author}
  {\bibfnamefont {T.~M.}\ \bibnamefont {Babinec}}, \bibinfo {author}
  {\bibfnamefont {A.}~\bibnamefont {Kubanek}}, \bibinfo {author} {\bibfnamefont
  {A.}~\bibnamefont {Yacoby}}, \bibinfo {author} {\bibfnamefont {M.~D.}\
  \bibnamefont {Lukin}}, \ and\ \bibinfo {author} {\bibfnamefont
  {M.}~\bibnamefont {Loncar}},\ }\href {\doibase 10.1021/nl204449n} {\bibfield
  {journal} {\bibinfo  {journal} {Nano Letters}\ }\textbf {\bibinfo {volume}
  {12}},\ \bibinfo {pages} {1578} (\bibinfo {year} {2012})}\BibitemShut
  {NoStop}%
\bibitem [{\citenamefont {Neu}\ \emph {et~al.}(2014)\citenamefont {Neu},
  \citenamefont {Appel}, \citenamefont {Ganzhorn}, \citenamefont
  {Miguel-Sanchez}, \citenamefont {Lesik}, \citenamefont {Mille}, \citenamefont
  {Jacques}, \citenamefont {Tallaire}, \citenamefont {Achard},\ and\
  \citenamefont {Maletinsky}}]{Neu2014}%
  \BibitemOpen
  \bibfield  {author} {\bibinfo {author} {\bibfnamefont {E.}~\bibnamefont
  {Neu}}, \bibinfo {author} {\bibfnamefont {P.}~\bibnamefont {Appel}}, \bibinfo
  {author} {\bibfnamefont {M.}~\bibnamefont {Ganzhorn}}, \bibinfo {author}
  {\bibfnamefont {J.}~\bibnamefont {Miguel-Sanchez}}, \bibinfo {author}
  {\bibfnamefont {M.}~\bibnamefont {Lesik}}, \bibinfo {author} {\bibfnamefont
  {V.}~\bibnamefont {Mille}}, \bibinfo {author} {\bibfnamefont
  {V.}~\bibnamefont {Jacques}}, \bibinfo {author} {\bibfnamefont
  {A.}~\bibnamefont {Tallaire}}, \bibinfo {author} {\bibfnamefont
  {J.}~\bibnamefont {Achard}}, \ and\ \bibinfo {author} {\bibfnamefont
  {P.}~\bibnamefont {Maletinsky}},\ }\href {\doibase
  http://dx.doi.org/10.1063/1.4871580} {\bibfield  {journal} {\bibinfo
  {journal} {Applied Physics Letters}\ }\textbf {\bibinfo {volume} {104}},\
  \bibinfo {eid} {153108} (\bibinfo {year} {2014})}\BibitemShut {NoStop}%
\end{thebibliography}
\end{document}